\documentclass[aps,pre,floatfix,longbibliography]{revtex4-1}
\usepackage[dvipsnames,rgb,dvips]{xcolor}
\usepackage{graphicx}
\usepackage{float}
\usepackage{overpic}
\usepackage{bbm}
\usepackage{amsmath,amssymb}
\usepackage{amsfonts}
\makeatletter
\def\amsbb{\use@mathgroup \M@U \symAMSb}
\makeatother
\usepackage[bbgreekl]{mathbbol}
\usepackage{units,mathrsfs}

\newcommand\nn{\nonumber}
\newcommand\ve[1]{\boldsymbol{#1}}
\newcommand{\ma}[1]{\ensuremath{\amsbb{#1}}}

\usepackage{array}
\usepackage[color=green!40]{todonotes}
\usepackage[normalem]{ulem}
\usepackage{textcomp}
\usepackage{gensymb} 
\newcolumntype{C}[1]{>{\centering\let\newline\\\arraybackslash\hspace{0pt}}p{#1}}

\newcommand{\lc}{\ensuremath{\varepsilon}}

\newcommand{\rd}{\ensuremath{\mathrm{d}}}
\newcommand{\Pe}{\ensuremath{\mathrm{Pe}}}
\newcommand{\invpe}{\ensuremath{\mathrm{Pe}^{-1}}}

\newcommand{\obs}[1]{{#1}}

\newcommand{\eqnlab}[1]{\label{eqn:#1}}
\newcommand{\figlab}[1]{\label{fig:#1}}
\newcommand{\eqnref}[1]{(\ref{eqn:#1})}
\newcommand{\figref}[1]{\ref{fig:#1}}
\newcommand{\Eqnref}[1]{Eq.~(\ref{eqn:#1})}
\newcommand{\Figref}[1]{Fig.~\ref{fig:#1}}
\newcommand{\Tabref}[1]{Table~\ref{tab:#1}}

\newcommand{\applab}[1]{\label{app:#1}}
\newcommand{\seclab}[1]{\label{sec:#1}}
\newcommand{\Appref}[1]{Appendix~\ref{app:#1}}
\newcommand{\Secref}[1]{Section~\ref{sec:#1}}
\newcommand{\St}{\ensuremath{\textrm{St}}}
\newcommand{\tr}{\ensuremath{^{\mathrm T}}}

\newcommand\hl{\bgroup\markoverwith %
  {\textcolor{yellow}{\rule[-.5ex]{2pt}{2.5ex}}}\ULon}

\begin{document}

\title{Intrinsic viscosity of a suspension of weakly Brownian ellipsoids in shear}
\author{G. Almondo$^{\ast 1}$, J. Einarsson$^{\ast 1}$, J. R. Angilella$^{2}$, B. Mehlig$^{1}$}
\affiliation{$^{1}$Department of Physics, University of Gothenburg, 41296 Gothenburg, Sweden.}
\affiliation{$^{2}$Universit\'e de Caen, Cherbourg, France.\\
\mbox{}$^\ast$\rm These authors contributed equally to this work.}

\date{\today}

\begin{abstract} 
We analyze the angular dynamics of triaxial ellipsoids in a shear flow subject to weak thermal noise. By numerically integrating an overdamped angular Langevin equation, we find the steady angular probability distribution for a range of triaxial particle shapes. From this distribution we compute the intrinsic viscosity of a dilute suspension of triaxial particles. We determine how the viscosity depends  on particle shape in the limit of weak thermal noise.  While the deterministic angular dynamics depends very sensitively on particle shape, we find that the shape dependence of the intrinsic viscosity is weaker, in general, and that   suspensions of rod-like particles are the most sensitive to breaking of axisymmetry. The intrinsic viscosity of a dilute suspension of triaxial particles is smaller than that of a suspension of axisymmetric particles
with the same volume, and the same ratio of major to minor axis lengths.
\end{abstract}
\maketitle

\section{Introduction}\label{sec:intro}
\citet{einsten1906,einstein1911} calculated the shear viscosity $\mu^*$ of a dilute suspension of non-interacting spheres in a viscous fluid. He found $\mu^*=\mu(1+\eta\phi)$, where $\mu$ is the viscosity of the suspending fluid, $\eta=5/2$ is the \emph{intrinsic viscosity}, and $\phi$ is the concentration by volume of the suspended spheres.
The suspension viscosity is larger than that of the suspending fluid
because the particle cannot deform as the suspension is sheared. There is extra stress in the particle to resist the surface traction from the flow, and therefore there is a contribution proportional to the volume fraction of particles \cite{Batchelor1970}.

For a non-spherical particle, this additional stress depends on the orientation of the particle relative to the shear flow, and it also depends upon the particle shape.
\citet{Jeffery1922} calculated the angular motion and dissipation for a small  ellipsoidal particle in order to determine the intrinsic viscosity $\eta$ for a dilute suspension of ellipsoids. He found that the angular motion, and consequently the intrinsic viscosity, depends indefinitely on the initial orientation of the ellipsoid. This indeterminacy is physically unsatisfactory because the macroscopic suspension viscosity $\mu^*$ should not depend on the detailed microscopic initial conditions of the suspended particles after a long time.

For larger particles, the effects of inertia may break this indeterminacy \cite{Jeffery1922,Saffman1956b,harper1968,Subramanian2005,einarsson2015a,einarsson2015b,rosen2015}. But the long-time dynamics still depends on the initial condition for sufficiently flat disk-shaped particles \cite{einarsson2015b}, which could lead to hysteresis in the rheological functions of an inertial suspension.

For small particles, thermal fluctuations render the particle trajectories stochastic, and eventually independent of their initial conditions. In this case the intrinsic viscosity $\eta$ is a function of particle shape and noise strength, when averaged over an ensemble of stochastic realizations \cite{brenner1970,Hinch1972}.
For spheroidal particles subject to sufficiently weak noise, the stationary angular distribution is independent of noise strength \cite{Leal1971,Hinch1972}. This is because the angular dynamics is well described by the deterministic Jeffery trajectories in this limit, but with occasional jumps to a nearby trajectory. After many such jumps, a stationary probability distribution over the deterministic trajectories is established, however, the time to reach equilibrium is longer for weaker noise strength.
The dilute, weak-noise rheology is given by averaging over this stationary distribution. The intrinsic viscosity of a suspension of spheroids is larger that that of a suspension of spheres, and the shape dependence is stronger for prolate spheroids than oblate spheroids \cite{Hinch1972}.

How do these results generalize to triaxial ellipsoids? Much less is known concerning particles that do not possess axisymmetry. In absence of noise, the angular trajectory of a triaxial ellipsoid in shear flow is doubly periodic or chaotic, but nevertheless depends indefinitely upon initial condition \cite{Gierszewski1978,Hinch1979,Yarin1997,ein2015exp}. Similarly to the case of axisymmetric ellipsoids, thermal fluctuations eventually establish a stationary distribution over these trajectories, and this angular distribution determines the suspension rheology. For strong noise \citet{Rallison1978} and \citet{Haber1984} determined the first deviations from the uniformly distributed equilibrium state. But the angular distributions and the resulting intrinsic viscosity in the weak noise regime remain unknown. It is hard to make analytical progress, because the deterministic dynamics is chaotic.

In this paper we numerically compute the angular distribution and resulting intrinsic viscosity for a range of triaxial ellipsoids in shear flow, subject to weak thermal noise. We derive the appropriate Langevin equation and solve it numerically for the stationary probability distribution. We show how the angular distribution reflects the underlying deterministic trajectories. We compute the resulting intrinsic viscosity for a dilute suspension and show that it is maximal for axisymmetric particle shapes. In general the shape dependence of the intrinsic viscosity is weaker than that of the deterministic angular dynamics, which depends very sensitively on particle shape.

The remainder of this paper is organized as follows.
In \Secref{theory} we present our notation, derive the Langevin equation, and give the relation between the angular distribution and the dilute suspension viscosity.
\Secref{results} contains the numerical results from our Langevin simulations. We discuss the results in \Secref{discussion} and conclude in \Secref{conclusions}.

\section{Theory}\seclab{theory}
\subsection{Notation}
Where possible we use vector notation without indices. We write vectors as $\ve a$, and their
components in the lab frame as $a_i$. Tensors are denoted by $\ma A$, and  $A_{ij\ldots}$ are the lab-frame components of this tensor. In some instances we find index notation necessary for clarity, and then we use the Einstein summation convention.
Contractions of adjacent indices are denoted by the dot product, as for example in the scalar product between two vectors $\ve a \cdot \ve b = a_i b_i$.
The double dot product denotes contraction of two adjacent indices. For example, $(\ma A:\ma B)_{il}=A_{ijk}B_{kjl}$ denotes a contraction between the
two rank-3 tensors $\ma A$ and $\ma B$. These conventions apply also to contractions between tensors of different ranks. 

\begin{figure}
\begin{overpic}{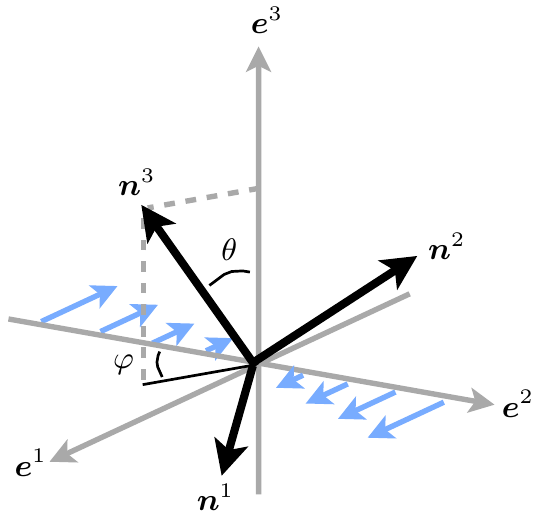}
\end{overpic}
\caption{\label{fig:shearplot} Coordinate system. Flow direction $\ve e^1$, 
shear direction $\ve e^2$. The flow vorticity points along $-\ve e^3$.
The angle between the principal  axis $\ve n^3$ and the $\ve e^3$-axis 
is $\theta$, and $\varphi$ is the angle between the $-\ve e^2$-axis
and the projection of $\ve n^3$ onto the flow-shear plane.
See appendix \ref{app:B} for details concerning the definition of the 
angles.}
\end{figure}

We represent the shape and orientation of an ellipsoid by the lengths ($a_1$, $a_2$, $a_3$) and directions $(\ve n^1, \ve n^2, \ve n^3)$ of its principal semi-axes. Without loss of generality we take $a_1\le a_2\le a_3$. The two aspect ratios are $\lambda=a_3/a_1$ and $\kappa=a_2/a_1$. 
We denote the coordinate axes of the lab frame by $(\ve e^1, \ve e^2, \ve e^3)$. They are  fixed with respect to the undisturbed fluid flow. The undisturbed flow takes
the form $\ve u^\infty = \ve \Omega^\infty \times \ve r+ \ma E^\infty\cdot \ve r$
where $\ve r$ is the spatial coordinate vector, $\ve \Omega^\infty$ is half the fluid vorticity, and $\ma E^\infty$ is the strain-rate matrix of the flow. We take the undisturbed flow to be a simple shear, $\ve u^\infty = (sr_2,0,0)$,  as shown in Fig.~\ref{fig:shearplot}. 

We also use the convention that components of a tensor in the particle coordinate frame have Greek indices, while components in the fixed lab frame have Latin indices, for example
\begin{align}
	\ve X &= \sum_{i=1}^3 X_i \ve e^i = \sum_{\alpha = 1}^3 X_\alpha \ve n^\alpha\,.
\end{align}
The two sets of components are related by the matrix $\ma R$,   defined by $\ve e^i = \ma R \ve n^i$, so that
\begin{align}
	X_\alpha &= R_{\alpha i} X_i\,,\quad X_i = R\tr_{i \alpha}X_\alpha\,.
\end{align}
Here $R\tr_{i\alpha}$ are the elements of $\ma R\tr$, the transpose of $\ma R$. Since the bases $\ve e^i$ and $\ve n^\alpha$ are both orthonormal, 
$\ma R$ is orthogonal, so that $\ma R^{-1} = \ma R\tr$. Appendix \ref{app:B}  explains how
the elements of $\ma R$ are expressed in terms of Euler angles \cite{Goldstein}.
The components of the  particle-orientation vector $\ve n^\beta$ in the lab frame are given by
\begin{align}
	n^\beta_i &= R\tr_{i\alpha}n^\beta_\alpha = R_{\beta i}\,.
\end{align}
In the remainder of this paper we employ dimensionless variables. We scale length by $V_p^{1/3}$ and time by $1/s$, where $V_p$ is the volume of the particle, and $s$ is the magnitude of the undisturbed shear rate. Stress is scaled by $\mu s$, where $\mu$ is the viscosity of the suspending fluid. 

\subsection{Orientational dynamics}
Disregarding thermal noise, the hydrodynamic angular velocity of an ellipsoidal particle in a linear Stokes flow $\ve u^\infty = \ve \Omega^\infty \times \ve r+ \ma E^\infty \cdot\ve r$ is given by
\begin{align}
	\ve \omega^\mathrm H &= \ve \Omega^\infty + \ma K^{-1}\cdot\,\ma H:\ma E^\infty\,.\eqnlab{omegaH}
\end{align}
The resistance tensors $\ma K$ and $\ma H$ are given by \citet{Haber1984} (see \Tabref{tableoftensors} for translation of notation). 
The tensor $\ma K$ describes the viscous resistance against a steady rotation of the particle, while $\ma H$ determines the effect of fluid strain on the hydrodynamic torque. The components of the resistance tensors are constant when expressed in the body frame, conversely the components of the flow gradients ($\ve \Omega^\infty$ and $ \ma E^\infty $)  are constant when expressed in the fixed lab frame. Therefore the components of the angular velocity, either in the body or the  lab frame, depend on the orientation of the particle.

Thermal fluctuations randomize the particle orientation.  The resulting angular probability distribution $P(\ma R, t)$ is governed by the Fokker-Planck equation
\cite{Rallison1978,Haber1984,bossis1989}
\begin{align}
	\frac{\partial P}{\partial t} &{+} \nabla \cdot \left(\ve \omega^{\rm H} {-} \Pe^{-1}\ma D \cdot \nabla\right){P}{=0}\,.\eqnlab{FPE}
\end{align}
Here $\Pe=\mu V_p s/(k_\mathrm B T)$ is the P\'eclet number which is a dimensionless measure of the noise strength. 
The \obs{non-dimensional} diffusion tensor is given by $\ma D = \ma K^{-1}$. A corresponding relation for the centre-of-mass diffusion of a small particle was first deduced by \citet{einsten1906,einstein1911}. In the steady state the diffusion flux must equal the flux due  to the external force, resulting in the relation \obs{$D = k_B T/(\mu V_p)$ for the dimensional rotational diffusion constant (which has the units of inverse of time).} Einstein's argument was adapted to the rotation of triaxial ellipsoids by \citet{Perrin1934}, and leads directly to 
Eq.~(\ref{eqn:FPE}): the first term in the parentheses on the l.h.s. of this equation is the angular flux due to the \obs{imposed flow, a shear in our case. The} second
term is the angular diffusion flux. Our notation is closest to that of \citet{Rallison1978} who studied this equation in the limit of strong noise. The gradients in the Fokker-Planck equation (\ref{eqn:FPE}) have the components 
$\nabla_k = \lc_{kij} R_{\alpha i} \partial_{\alpha j}$ where $\partial_{\alpha j}$ is the unconstrained differential in the nine-dimensional space of linear transformations \cite{Rallison1978}, and
 $\lc_{kij}$ is the Levi-Civita symbol denoting the elements of the completely antisymmetric third-rank tensor.

We do not solve Eq.~(\ref{eqn:FPE}) directly. Instead we consider
the equivalent Langevin equation \cite{VanKampen1981,bossis1989} for the angular increments
$\delta \ma R$ during the time interval $\delta t$:
\begin{subequations}\eqnlab{langevindef}
\begin{align}
	&\ma R(t+\delta t) = \ma R(t) + \delta \ma R(t)\,, \eqnlab{langevindefa}\\
	&\overline{\delta R_{\alpha i}} = -\lc_{\alpha\beta\gamma}\omega^{\mathrm H}_\beta(t) R_{\gamma i}(t) \delta t + \invpe \lc_{\alpha\beta\gamma} \lc_{\gamma \rho \sigma} K^{-1}_{\beta \rho}R_{\sigma i}(t)\delta t\,,\eqnlab{langevindefb} \\	&\overline{\delta R_{\alpha i}\delta R_{\mu p}} = 2\invpe \lc_{\alpha \beta \gamma}\lc_{\mu \rho \sigma}R_{\gamma i}(t)R_{\sigma p}(t)K^{-1}_{\beta \rho}\delta t\,.\eqnlab{langevindefc}
\end{align}
\end{subequations}
In the limit of weak noise, Eqs.~(\ref{eqn:langevindef}) are valid provided that $\St \ll \delta t  \ll 1$.  The lower limit for $\delta t$ is given by the Stokes number $\St=\rho_p s V_p^{2/3}/\mu$, a dimensionless measure of the particle inertia, where $\rho_p$ is the particle density. We must assume that the Stokes number is small enough so that the condition on $\delta t$ can be satisfied.
In this paper we consider the limit of large (but finite) P\'eclet numbers. In this case the upper limit for $\delta t$ is determined by the shear rate, so it is equal to unity in our dimensionless variables.
In \Eqnref{langevindefb} and \eqnref{langevindefc} the over-bar denotes an average over fluctuating angular displacements at fixed initial particle orientations, distinct from the thermal average $\langle\cdots\rangle$ over the steady-state distribution of orientations. 

\obs{The Langevin equation (\ref{eqn:langevindef}) can be derived directly from the angular-momentum equation \cite{bossis1989}
\begin{align}
	&\St \frac{\rd (\ma I\ve \omega)}{\rd t} = \ma K\cdot (\ve \Omega^\infty-\ve \omega) + \ma H : \ma E^\infty + \ve \Gamma(t)\,,\eqnlab{torquedimless}\\
	&\langle\ve \Gamma(t)\rangle=0\,,\quad\langle\ve \Gamma(t_1)\ve \Gamma\tr(t_2)\rangle=2\invpe\ma K \,\delta(t_1-t_2)\,.\nn	
\end{align}
Here $\ma I$ is the moment of inertia of the particle, and
the stochastic torque $\ve \Gamma(t)$ represents the torque due to thermal motion in the fluid. The angular brackets $\langle\cdots\rangle$ denote an average over thermal noise. The random torque has a very short (molecular) correlation
time $\tau$, represented by the correlation function $\delta_\tau(t)$ in \Eqnref{torquedimless}, and its statistics are determined by equipartition: the particle must be in thermal equilibrium with the surrounding fluid \cite{Perrin1934}. One integrates \Eqnref{torquedimless} 
for a small time step $\delta t$, together with the} kinematic equation 
\begin{align}
\tfrac{\rd }{\rd t}R_{\alpha i} = \lc_{ijk}\omega_jR_{\alpha k} = -\lc_{\alpha\beta\gamma}\omega_\beta R_{\gamma i}&\eqnlab{Rmotion}
\end{align}
\obs{that describes the rotation of the particle-orientation vectors $\ve n^\beta$ with angular velocity $\ve\omega$.
 The small time step  is assumed to be much smaller than the time over which the distribution
 of $\ma R(t)$ changes  ($\delta t\ll 1$ for large ${\rm Pe}$), yet large compared
 to the viscous time ($\delta t \gg {\rm St}$). In this limit one finds the following expressions for the moments of the particle
angular velocity \cite{Perrin1934}: }
\begin{subequations}
\begin{align}
\label{eq:meanomega}
	&\int\limits_0^{\delta t}\rd t \,\langle \ve \omega (t) \rangle \sim \ve \omega^{\rm H} \delta t\quad { \mbox{for}\quad
	\St \ll \delta t\ll 1},
	\end{align}
	and
	\begin{align}
	&\int\limits_0^{\delta t}\rd t_1\int\limits_0^{\delta t}\rd t_2\, \langle (\ve \omega(t_1)-\ve\omega^{\rm H}) \cdot(\ve \omega(t_2)-\ve\omega^{\rm H})\rangle \sim 2\invpe\ma K^{-1} \delta t\quad { \mbox{for}\quad
	\St \ll \delta t\ll 1}\,.\eqnlab{omegacorrelation}
\end{align}
\end{subequations}
The asymptotic form of the integral of the angular-velocity autocorrelation function  is consistent with Einstein's argument mentioned above. We note that the minus sign in Eq.~(\ref{eqn:Rmotion}) arises from transforming the equation of motion to body coordinates. This explains the minus sign in Eq.~(\ref{eqn:langevindefb}), written in body coordinates.
Finally we remark that the second term on the r.h.s. of Eq.~(\ref{eqn:langevindefb}) is a spurious drift term \cite{VanKampen1981}. It arises here because $\tfrac{{\rm d}}{{\rm d}t}{\ma R}$ is a non-linear function of \ma R since $\ve \omega^{\rm H}$ depends upon the particle orientation through the resistance tensors, and thus on $\ma R$. An analogous situation is described in Ref.~\cite{Meh05}. \obs{Finally, averaging the angular displacements $\delta \ma R$ at fixed initial orientation one finds, using (\ref{eq:meanomega}) and (\ref{eqn:omegacorrelation}), the Langevin equation (\ref{eqn:langevindef})}.

To simulate \Eqnref{langevindef} in practice, we represent the orientation by a unit quaternion instead of a rotation matrix \cite{graf2008}. The unit quaternion is better than the rotation matrix for numerical computation because it has four scalar components and a unit constraint $|\ve q|=1$, whereas the rotation matrix has nine scalar components and the orthogonality constraint $\ma R\tr\ma \cdot \ma R=\mathbbm{1}$. The Langevin equation in quaternion coordinates is described in \Appref{quaternion}.
\begin{table}
\caption{\label{tab:tableoftensors} The dimensionless elements of the resistance tensors in relation to expressions in \citet{Haber1984}. \obs{A factor of $2$ is missing in Eq.~[A1] for $^r\ve{\mathrm K}$ on p. 510 of \citet{Haber1984}.
We have used the correct expression from Ref.~\cite{Kim1991}.}}
\begin{ruledtabular}
\begin{tabular}{cll}
Notation in this paper & Notation in Ref.~\cite{Haber1984} & Eqs. in Ref. \cite{Haber1984}\\
\hline
$\ma K$ & $6\,^r\ve{\mathrm K}$ & Eqs.~[3.1], [A1]\\
$\ma H$ & $6\ve \tau$ & Eqs.~[3.1], [A2]\\
$\ma Z$ & $5\ve{\mathrm Q}$ & Eqs.~[3.1], [A3], [A4] 
\end{tabular}
\end{ruledtabular}
\end{table}
\subsection{Dilute suspension rheology}
The macroscopic description of a particulate suspension is based on a statistical model of the microscopic fluid mechanics of all the suspended particles \cite{brenner1972}. For a sufficiently homogeneous suspension, a macroscopic observable such as the stress tensor $\bbsigma$, may be represented by an average of the microscopic configurations. In general this averaging is a very complicated task \cite{Batchelor1970}. But for a dilute suspension it is sufficient to consider the stress contribution from an isolated particle and sum the independent contributions from all particles, because particle interactions are negligible. This gives the correct rheology to first order in the volume fraction of particles in the suspension \cite{Batchelor1970,brenner1972}.

\citet{Batchelor1970} showed that the stress contribution from a 
single torque-free particle in steady Stokes flow is determined by the symmetric force dipole on the particle, the so-called stresslet. In terms of resistance tensors, the stresslet  for a torque-free particle is
\begin{align}
	\ma S &= \ma C : \ma E^\infty\,,\eqnlab{avgstressletmain}
\end{align}
where the components of $\ma C$ are 
\begin{align}
	C_{ijkl} &= R_{\alpha i}R_{\beta j}R_{\gamma k}R_{\delta l}\left(Z_{\alpha \beta \gamma \delta}-H_{\mu \alpha \beta}K^{-1}_{\mu \nu}H_{\nu \gamma \delta}\right)\,.\eqnlab{Ctensor}
\end{align}
The rank-four tensor $\ma Z$ is the resistance tensor coupling stresslet and strain (\Tabref{tableoftensors}).
\Eqnref{avgstressletmain} was derived by \citet{Batchelor1970} in the steady Stokes approximation, assuming
 weak thermal noise (large P\'eclet numbers). While the Langevin equation (\ref{eqn:langevindef})
is valid for arbitrary P\'eclet numbers, {\em unsteady} \obs{fluid inertia}  might affect the Brownian contribution to the stress for triaxial particles.
 It may well be, on the other hand, that the steady Stokes approximation is sufficient. To show this one should find an argument -- analogous to Einstein's -- that shows that the steady Stokes approximation gives the correct result.  
  \obs{For a diffusing sphere it is known that the velocity autocorrelation function is wrongly predicted by the steady Stokes approximation, yet the long-time mean squared displacement of the centre-of-mass comes out correctly.}

Here we consider the limit of   large P\'eclet numbers, to avoid this question.
In this limit the extra stress $\mathbb{\Sigma}_{{p}}$
due to the presence of particles in a dilute suspension of volume $V$ is given by \cite{Batchelor1970}
\begin{align}
	\mathbb{\Sigma}_{{p}} = \frac{V_p}{V} \sum_{m}{\ma S}^{{{(m)}}}\,,
	\label{Sigmap}
\end{align}
where ${\ma S}^{{{(m)}}}$ denotes the stresslet from the $m$:th particle and the sum is over all particles. The stresslet \eqnref{avgstressletmain} depends on the the shape and orientation of the particle. If there are many identical particles in $V$,	 the sum over particles may be replaced by an angular average over the distribution $P(\ma R)$:
\begin{align}
	\mathbb{\Sigma}_{{p}} = \phi 
\int \rd\ma R\,\,{\ma S}(\ma R)\,P(\ma R)\,.\eqnlab{orientationalavg}
\end{align}
Here $\phi=N V_p/V$ is the volume concentration of particles, with $N$ 
the number of particles in the volume $V$. The volume concentration is assumed
to be small, $\phi\ll 1$.
The intrinsic viscosity $\eta$ is determined by the element
 $\Sigma_{{p};12}$, 
the shear stress due to the particles  \cite{giesekus1962,Rallison1978}:
\begin{equation}
\label{eqn:iv}
\eta \equiv  \frac{1}{\phi}\Big(\frac{\mu^\ast}{\mu}-1\Big)
 = \frac{ \Sigma_{{p};12}}{\phi}\,.
\end{equation}
This rheological function depends on the particle shape ($\lambda,\kappa$)
and on the value of $\Pe$. When thermal noise is significant there are extra direct Brownian contributions to the stress that we have not considered here. Therefore
we only consider the limit of weak thermal noise, corresponding to
large values of the P\'eclet number. In this limit we expect the angular distribution and therefore the intrinsic viscosity to converge to $\Pe$-independent values \cite{Leal1971,Hinch1972}, so that the viscosity becomes independent
of ${\rm Pe}$, and direct contributions to the stress from Brownian rotation are negligible.

\begin{figure}
\begin{overpic}[width=8cm]{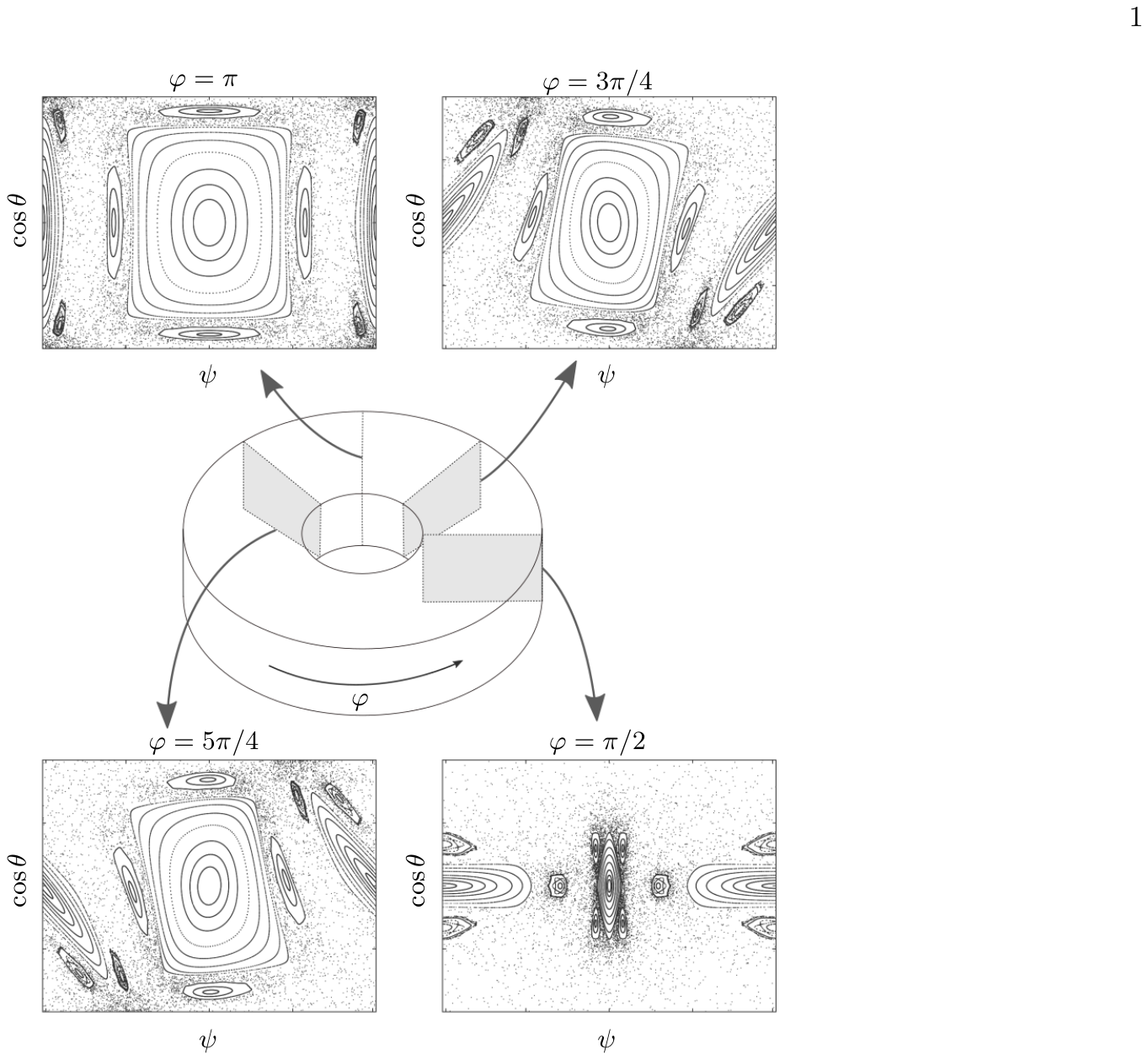}
\end{overpic}
\caption{\label{fig:torus} Schematic explanation of the surfaces-of-section 
shown in \Secref{distributions}  (for $\lambda=10$ and $\kappa=5$). The torus depicts the three-dimensional angular space for an ellipsoid. The major axis of the ellipsoid rotates monotonously around the vorticity \cite{Hinch1979}, depicted by the azimuthal angle $\varphi$,
see also Fig.~\ref{fig:shearplot}. The surfaces-of-sections shown are for 
$\varphi\!=\pi/2+\!n\pi/4$, for $n\!=\!0,\ldots,3$, 
corresponding to four directions of the projection of $\ve n^3$ to the flow shear plane: parallel with the flow,  of extending strain, 
perpendicular to the flow, and of compressing strain.  The surfaces-of-section for $\varphi \to \varphi+\pi$ are equal, because the problem is symmetric under this rotation. Figure reproduced from Ref.~\citenum{einarssonphd} under the CC-BY 3.0 license.}
\end{figure}

\begin{figure*}
\begin{overpic}{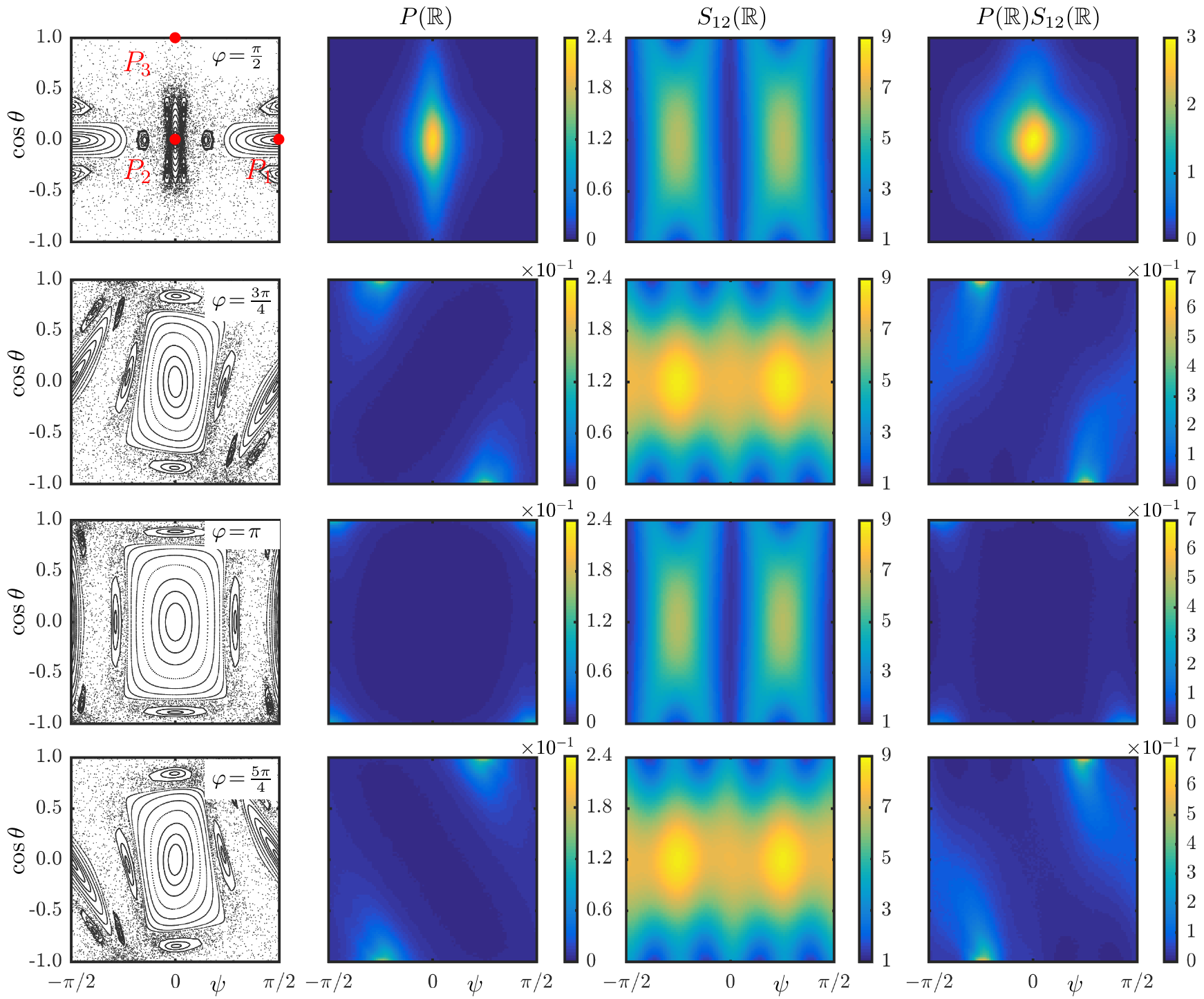}
\end{overpic}
\caption{\label{fig:distlownoise_k5} 
Rows: four representative surfaces-of-sections for $\varphi\!=\pi/2+\!n\pi/4$, 
for $n\!=\!0,\ldots,3$. These sections correspond to four directions of the projection of $\ve n^3$ to the flow shear plane: parallel with the flow, of extending strain, perpendicular to the flow, and of compressing strain (Fig.~\ref{fig:torus}).
Columns: \emph{(1)} the surface-of-section of deterministic trajectories; \emph{(2)} the stationary angular distributions; \emph{(3)} the stresslet element corresponding to the intrinsic viscosity of a dilute suspension; \emph{(4)} the contribution to intrinsic viscosity, given by the product of the angular  distribution and the stresslet element. Parameters: $\lambda=10$, $\kappa=5$, and $\Pe=200$. The points $P_i$ indicate the locations of three periodic orbits with $\ve n^i$ parallel to vorticity, 
for $i=1,\ldots,3$ (see text).} 
\end{figure*}

\begin{figure*}
\begin{overpic}{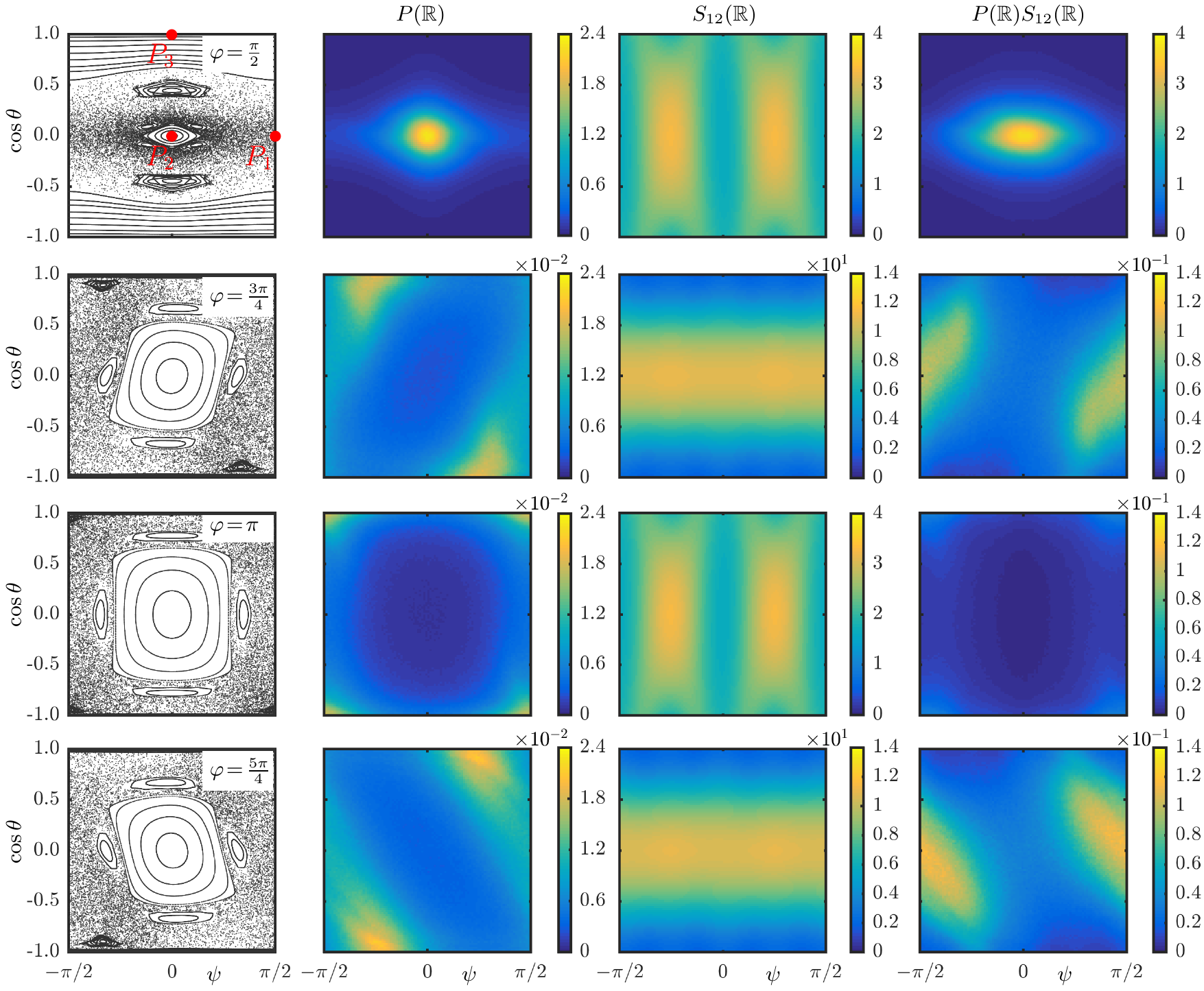}
\end{overpic}
\caption{\label{fig:distlownoise_k2} 
Rows: four representative surfaces-of-sections for 
$\varphi\!=\pi/2+\!n\pi/4$, $n\!=\!0..3$, 
corresponding  to four directions of the projection of $\ve n^3$ to the flow shear plane: parallel with the flow, of extending strain,
perpendicular to the flow, and of compressing strain (Fig.~\ref{fig:torus}).
Columns: \emph{(1)} the surface-of-section of deterministic trajectories; \emph{(2)} the stationary angular distributions; \emph{(3)} the stresslet element corresponding to the intrinsic viscosity of a dilute suspension; \emph{(4)} the contribution to intrinsic viscosity, given by the product of the angular distribution and the stresslet element. Parameters: $\lambda=10$, $\kappa=2$, and $\Pe=200$. The points $P_i$ indicate the locations of three periodic orbits with $\ve n^i$ parallel to vorticity, 
for $i=1,\ldots,3$ (see text).} 
\end{figure*}

\section{Numerical results}\seclab{results}

In this Section we show numerical results for the stationary angular distribution and the resulting intrinsic viscosity.

\subsection{Orientational distributions}\seclab{distributions}
The angular trajectories and distributions of a rigid body are difficult to visualize, because they are defined in the three-dimensional non-Euclidean orientation space. But in a particular Euler-angle representation $(\varphi, \theta, \psi)$ (see Appendix B) \citet{Hinch1979} found that $\dot \varphi < 0$ in absence of noise. This means that the particle monotonously rotates around the vorticity axis.  Therefore it is helpful to think about the orientation space as a torus, in which the deterministic trajectories go around, see \Figref{torus}. Each transversal slice of constant $\varphi$ of this torus is a Poincar\'e surface-of-section \cite{Strogatz}, schematically shown in \Figref{torus}. 
To illustrate the angular distributions, we choose the four representative surfaces-of-sections 
for $\varphi\!=\pi/2+\!n\pi/4$, for $n\!=\!0,\ldots,3$, 
corresponding to four directions of the projection of $\ve n^3$ to the flow shear plane: parallel with the flow, of extending strain,
perpendicular to the flow, and of compressing strain (Fig.~\ref{fig:torus}).
 The first columns of Figs.~\figref{distlownoise_k5} and \figref{distlownoise_k2} show these four surfaces of section
for two particle shapes:  \Figref{distlownoise_k5} is for a strongly triaxial ellipsoid with aspect ratios $\lambda=10$ and $\kappa=5$, while \Figref{distlownoise_k2} is for a moderately triaxial particle with $\lambda=10$ and $\kappa=2$.
To obtain these plots we simulated the deterministic angular
dynamics (at ${\rm Pe}=\infty$) and verified that the integration step size $\delta t$ was small enough not to affect the results shown.
  We note that the surfaces of Section in Refs.~\cite{Hinch1979,Yarin1997,ein2015exp}
are for $\varphi = n \pi$ ($\ve n^3$ perpendicular to the flow).

Another way of visualising the deterministic angular dynamics is to analyse its periodic solutions. 
\citet{Yarin1997} described three periodic orbits that correspond to the rotation of the triaxial ellipsoid around
$\ve n^1$, $\ve n^2$, and $\ve n^3$. The points $P_i$ in Figs. \ref{fig:distlownoise_k5} and \ref{fig:distlownoise_k2} indicate where these periodic orbits ($\ve n^i$ parallel to vorticity)  intersect the surfaces-of-section.

Now consider the stochastic angular dynamics. We show distributions of $\ma R$ for $\lambda=10$
and $\kappa=2,5$ at weak noise ($\Pe=200$). We verified our numerical 
algorithm for computing these distributions by evaluating different moments for axisymmetric particles, and found them to be in good agreement with the results
 of Ref.~\cite{Hinch1972,Rallison1978} and \cite{Ein14}.
This criterion does not test the far tails of the distributions which are difficult to calculate with high accuracy at large P\'eclet numbers. Therefore we chose a relatively small value of Pe here, about 10 times smaller than
the values used for calculating the intrinsic viscosity (Section \ref{sec:iv}).

The second columns in Figs.~\figref{distlownoise_k5} and \figref{distlownoise_k2} show our results for the stationary angular distributions. The third columns show the stresslet element corresponding to the intrinsic viscosity of a dilute suspension. In the fourth columns we plot the contribution to intrinsic viscosity, given by the product of the angular distribution and the stresslet element. 
At low thermal noise the distribution is dominated by the deterministic dynamics, 
and the only effect of the noise is to establish a distribution over the 
deterministic trajectories.  For strong thermal noise, by contrast,
the distribution $P(\ma R)$ is nearly isotropic (not shown).

\subsection{Intrinsic viscosity}
\label{sec:iv}
\begin{figure*}
\begin{overpic}{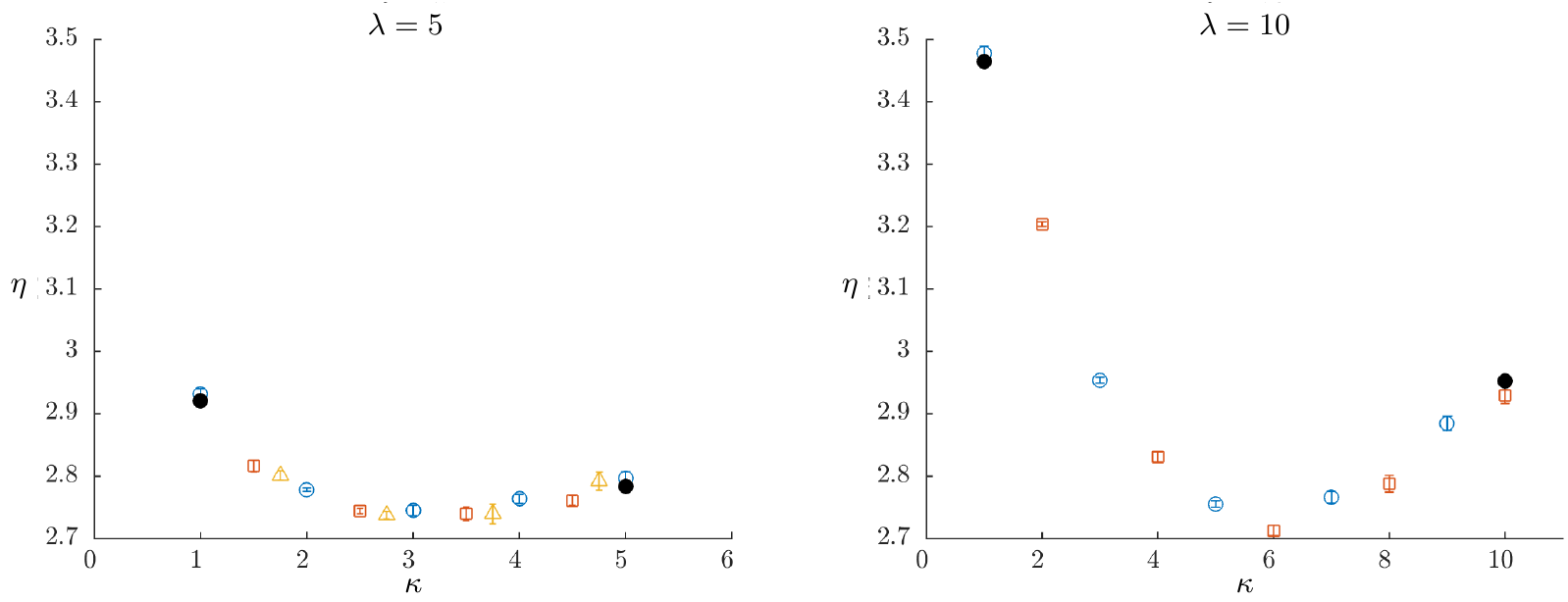}
\end{overpic}
\caption{\figlab{viscvsk} Intrinsic viscosity of a dilute suspension of triaxial ellipsoids as a function of $\kappa$. Left $\lambda=5$, 
$\Pe=500$ (\textcolor{blue}{$\circ$}), $\Pe=1000$ (\textcolor{red}{$\Box$}),
 $\Pe=2000$ (\textcolor{yellow}{$\bigtriangleup$}). Right: $\lambda=10$, $\Pe=2000$ (\textcolor{blue}{$\circ$}), $\Pe=3000$ (\textcolor{red}{$\Box$}). The error bars represent one standard deviation in our average over $200$ independent realisations of the Langevin process (\ref{eqn:langevindef}).
On the $x$-axis, the shapes represented by the $\kappa$-values 
range from rods ($\kappa=1$) to disks ($\kappa=\lambda$). 
Known limiting values for axisymmetric particles ($\bullet$)
are taken from Refs.~\cite{Leal1971,Hinch1972}.
}
\end{figure*}
From \Eqnref{iv} we computed the intrinsic viscosity in the limit of weak thermal noise, at large P\'eclet numbers. We chose $\Pe$ as large as practically possible, so that the intrinsic viscosity converges to a $\Pe$-independent plateau, as in the axisymmetric case \cite{Leal1971,Hinch1972}. 
\Figref{viscvsk} shows the results for spheroidal particles, 
for $\lambda=5$ and $10$ as a function of $\kappa$.
For the data shown we simulated $N=200$ independent instances of the Langevin equation (\ref{eq:ql}) for $5\cdot10^5$ \obs{dimensionless} time units, with timesteps $\delta t$ between $10^{-4}$ and $10^{-5}$.  The error bars  in \Figref{viscvsk} represent one standard deviation in our average over the $200$ independent realisations.

In \Figref{viscvsk}, the parameter $\kappa$ ranges from $\kappa =1$ to $\kappa=\lambda$. 
The limiting cases correspond to rotationally symmetric, ellipsoidal particles.
In these special cases our numerical results agree with those of previous work.
The values for $\lambda=10$ and $\kappa = 1,10$ are determined from Eq.~(27)
in \cite{Leal1971}, together with the angular averages 
from Table~1 in this paper. The angular averages for  $\lambda=5$ 
and $\kappa = 1,5$ are taken from Table~3 in \cite{Hinch1972}.
For the $\lambda\!=\!10$-particle slightly higher $\Pe$-values are needed to obtain this convergence than for the  $\lambda\!=\!5$-particle. 
We observe good agreement with these results for axisymmetric particles
(to within a fraction of a percent), but the agreement is not perfect. We checked that the remaining error
is not due to the finite integration step size $\delta t$ by varying this
step size. A possible source of error is  the statistical error due
to finite sample size, and we cannot rule out that the initial transient
may result in a small systematic error.

\section{Discussion}\seclab{discussion}
\subsection{Orientational distributions}

The deterministic angular trajectories depend very sensitively on the shape of the particle. While axisymmetric ellipsoids tumble on periodic Jeffery orbits, a slight breaking of this symmetry can lead to doubly periodic, and even chaotic angular dynamics \cite{Hinch1979,Yarin1997,ein2015exp}, as the the surfaces-of section in the first rows of Figs.~\ref{fig:distlownoise_k5} and \ref{fig:distlownoise_k2} show. 
The closed concentric lines near $\cos\theta=0$ in these surfaces-of-section  describe doubly-periodic tumbling,
\obs{while the black regions correspond to chaotic tumbling.} \obs{The} almost horizontal lines
near $|\cos\theta|=1$ correspond to slightly perturbed Jeffery orbits (log rolling). 

\obs{The} surfaces-of-section look very similar to those of Hamiltonian dynamics \cite{Strogatz}. This may appear surprising, because our dynamics is dissipative, not Hamiltonian. But it is no coincidence that the surfaces-of-section look so similar.  While our system does not conserve energy,
it is time-reversal invariant and exhibits \obs{a} discrete reflection symmetry \cite{ein2015exp}
that constrains the angular dynamics in a way analogous to the symplectic structure
of Hamiltonian dynamics \obs{\cite{Strogatz,Politi1986}}.  

For weak noise, the  particle orientation tends to follow deterministic trajectories, but occasionally jumps to a neighboring trajectory. This process establishes an equilibrium distribution of the particle orientation over the deterministic trajectories after some time.  Which orientations are most probable, and how does the distribution reflect the nature of the deterministic angular dynamics? 

Figs.~\figref{distlownoise_k5} and \figref{distlownoise_k2} show that the probability is highest in the flow-shear plane, when $\ve n^3$ aligns with the flow direction (first rows of Figs.~\figref{distlownoise_k5} and \figref{distlownoise_k2}). This is simply a consequence of the time-scale separation in the deterministic dynamics when $\lambda$ is not near $1$: elongated particles spend most of their time aligned with the flow where the angular dynamics is slow. This orientation corresponds to a local minimum of shear stress (third panel in first row of Figs.~\figref{distlownoise_k5} and \figref{distlownoise_k2}).

The other three surfaces of section capture how the angular dynamics when the projection of $\ve n^3$ is not aligned with the flow direction. The probability is not uniformly distributed over the surfaces of section. Also in this case peaks in $P(\ma R)$ are explained by slow angular dynamics. Consider the second row of Figs.~\figref{distlownoise_k5} and \figref{distlownoise_k2}, corresponding to $\varphi = 3\pi/4$. 
The probability is peaked at $(\cos\theta,\psi) \approx (1,-\pi/4)$ and
the symmetric point $(-1,\pi/4)$. The condition $\cos\theta=\pm 1$ corresponds
to the log-rolling orbit, and when $\cos\theta= 1$ then $\psi = -\pi/4$ ensures that the short axis $\ve n^1$ aligns with the shear direction where the shear-induced torque $\ma H : \ma E^\infty$ is minimal, so  that the angular dynamics is slow. The same argument holds for $\cos\theta= -1$ then $\psi = \pi/4$.
In rows $3$ and $4$ of Figs.~\figref{distlownoise_k5} and \figref{distlownoise_k2} 
the situation is analogous: the probability $P(\ma R)$ is largest for orientations where the shear-induced torque is smallest. 
Comparing Figs.~\figref{distlownoise_k5} and \figref{distlownoise_k2}
we see that the maximal values of $P(\ma R)$ are similar (first rows). This is expected because the parameter $\lambda$ is the same. The probability
in rows 2, 3, and 4 is larger in Fig.~\figref{distlownoise_k5} 
($\kappa = 5$) compared with Fig. \figref{distlownoise_k2} ($\kappa = 2$).
A larger value of $\kappa$ corresponds to slower dynamics, and thus to higher probability.
In summary, the probability $P(\ma R)$ of orientations in the weak-noise limit 
is strongly peaked where the deterministic dynamics is slowest, regardless of whether
it is  periodic, doubly periodic or possibly chaotic.

\subsection{Intrinsic viscosity}

The orientation-dependent contribution to intrinsic viscosity $S_{12}(\ma R)$, however, has a local minimum where the probability density is concentrated 
(Figs.~\figref{distlownoise_k5} and \figref{distlownoise_k2}).
Nevertheless, this direction dominates the contribution to the intrinsic viscosity at weak noise.
With the major axis along the flow direction, the orientation corresponding to maximal shear stress is when the particle is tilted $45\degree$.  
Although those particle orientations are relatively unlikely, they contribute to the integral of $P(\ma R)S_{12}(\ma R)$ because of their relatively high 
shear stress.

\Figref{viscvsk} shows the resulting intrinsic viscosity. We see that $\Pe$ is large enough so that the intrinsic viscosity is approximately independent of $\Pe$, to within numerical accuracy.  For the $\lambda\!=\!10$-particle slightly higher $\Pe$-values are needed to obtain this convergence, than for the  $\lambda\!=\!5$-particle. We believe this is because the effective $\Pe$ in regions of slow deterministic dynamics is actually smaller than the naive estimate $\Pe=s/D$ \obs{and} the more elongated \obs{the} particle \obs{is}, the slower \obs{is the} dynamics \obs{in such regions} \cite{Hinch1972}.

\begin{figure}[t]
\begin{overpic}[width=0.7\textwidth]{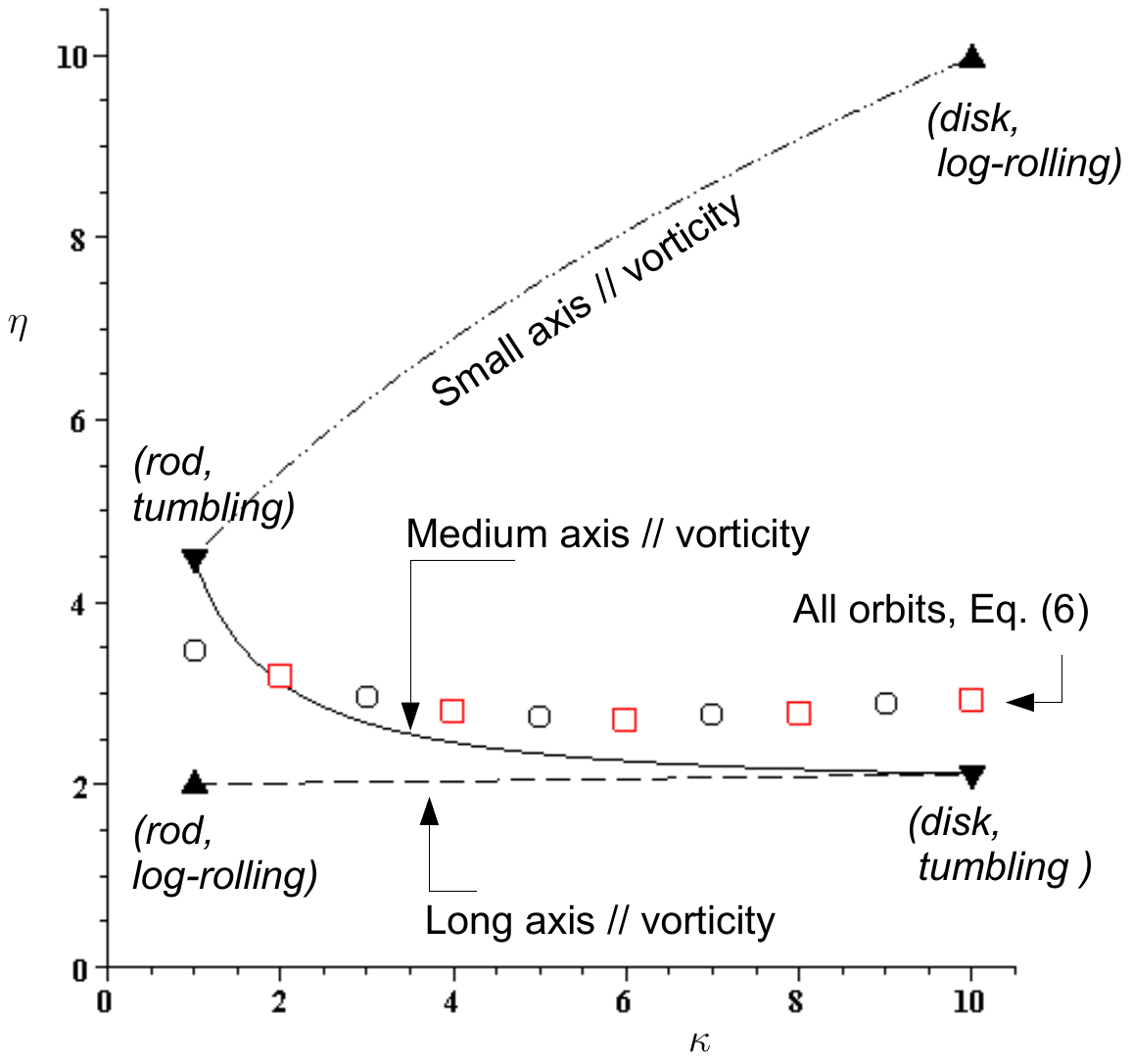}
\end{overpic}
\caption{\figlab{eta_periodic_orbits}  
Plot of the  intrinsic viscosities for the three periodic orbits corresponding
to rotation with one axis parallel to vorticity
(see text), as a function of $\kappa$ for  $\lambda = 10$. Solid line: intermediate axis parallel to vorticity, dashed line: long axis parallel to vorticity, dot-dashed line: short axis parallel to vorticity.
Triangles  correspond to values for axisymmetric particle that were computed by  \citet{Jeffery1922}:
tumbling ($\blacktriangledown$) and log rolling ($\blacktriangle$). Squares and circles are the numerical data  from Fig.~\ref{fig:viscvsk} (right),  for P\'eclet number equal to 2000 (circles) and 3000 (squares).}
\end{figure}

For all particle shapes shown the intrinsic viscosity is larger than that of spheres ($\eta=5/2$).
This is consistent with the observation that the intrinsic viscosity of a dilute suspensions of axisymmetric particles increases with larger particle aspect ratio \cite{Hinch1972}, most strongly for suspensions of prolate particles. 
The effect of making the particles triaxial, however, is to decrease the resulting intrinsic viscosity,
as can be seen in \Figref{viscvsk}. The Figure shows that the intrinsic viscosity depends
only  weakly on $\kappa$, except for rod-like axisymmetric particles. 
We conclude that the intrinsic viscosity does not depend as sensitively on particle
shape as the deterministic angular dynamics, even at low thermal noise where the angular dynamics
follows deterministic trajectories for long times. Figs.~\figref{distlownoise_k5} and \figref{distlownoise_k2} show that this is the consequence of two effects. First, the angular dynamics is most sensitive to particle shape near orientations where the particle spends least time. Second, the additional stress caused by the particle is comparatively small at these orientations.

To illustrate these conclusions in a different way, we computed the 
\obs{intrinsic} viscosities associated with 
the three periodic orbits mentioned above, where the triaxial ellipsoid rotates about one of its major
axes, $\ve n^i$ for $i=1,\ldots,3$. The first orbit ($P_1$ in Figs.~\figref{distlownoise_k5} and \figref{distlownoise_k2}) corresponds to a particle rotating with its small axis parallel to the undisturbed vorticity $\ve \Omega^\infty$, $P_2$ to rotation with the intermediate axis parallel to vorticity, and $P_3$ to rotation with the long axis parallel to vorticity. For each orbit we averaged
$S_{12}$ [Eq.~(\ref{eqn:iv})] along the orbit. The results are shown in Fig.~\figref{eta_periodic_orbits}.
We see that the orbit corresponding to  $P_1$ is the most dissipative one:  if all particles rotated with their small axis parallel to vorticity, the intrinsic viscosity would reach large values. However, this orbit is known to be unstable \cite{Yarin1997}, and has a low probability in the weak-noise dynamics (see Figs.~\ref{fig:distlownoise_k5} and \ref{fig:distlownoise_k2}).
The orbit corresponding to $P_2$  (medium axis parallel to vorticity), by contrast, has the highest probability of the three periodic orbits in the weak-noise dynamics, as Figs.~\ref{fig:distlownoise_k5} and \ref{fig:distlownoise_k2} show. It is much less dissipative though. This suggests that particles spend a long time with medium axis aligned with vorticity
in the weak-noise limit, and that this yields the dominant contribution to the overall intrinsic viscosity  of the dilute suspension.

\section{Conclusions}\seclab{conclusions}
We analyzed the angular dynamics of triaxial ellipsoids in a shear flow 
subject to weak thermal noise (large P\'eclet numbers). 
By numerically integrating the corresponding angular Langevin equation, we found the stationary probability distribution for a range of asymmetric particle shapes at weak thermal noise. We showed that the probability is largest when the deterministic
angular dynamics is slow, regardless of whether it is strictly periodic, doubly periodic, or chaotic. 

We also compared how the angular distribution correlates with the orientation-dependent contribution to the intrinsic viscosity of a dilute suspension. We found that the 
angular probability is concentrated in a local minimum of the shear stress.  In general though the shear stress is much less localized than the angular probability.

Finally, we computed the intrinsic viscosity of a dilute suspension of triaxial ellipsoids at weak noise, and found that the intrinsic viscosity decreases as particles deviate from axisymmetric shape \obs{(for particles with the same volume, and with the same ratio of major to minor axis lengths). This effect is strongest} for rod-shaped particles, \obs{it} is thus
important to ensure that rod-like particles  are axisymmetric to high precision when trying to achieve a maximal increase in suspension viscosity by adding rod-like particles to a suspension.
For example, at $\lambda=10$ changing $\kappa$ from $1$ to $2$ gives a $10\%$-reduction in intrinsic viscosity.  In general, however, we found that the dependence of the intrinsic viscosity on particle shape is much less sensitive than the nature of the deterministic angular dynamics, because the angular probability is localized where the shear-induced  torque is small, regardless of the nature of the classical dynamics. 

\obs{Finally, at lower values of $\mbox{Pe}$, suspensions of spheroids exhibit normal stress differences \cite{Hinch1972} 
that are $O(\mbox{Pe}^{-1})$ smaller than the shear-stress correction, and therefore outside the scope of the present study. Computing the rheological properties of suspensions of triaxial particles at lower values of Pe is an interesting future research opportunity.}

\acknowledgements{We thank K. Kroy for discussions.
We acknowledge  support by Vetenskapsr\aa{}det [grant numbers 2013-3992 and 2017-3865], 
by the grant \lq Bottlenecks for particle growth in turbulent aerosols\rq{} from the Knut and Alice Wallenberg Foundation, Dnr. KAW 2014.0048, and by the MPNS COST Action MP1305 \lq Flowing matter\rq{}.
The numerical computations used resources provided by C3SE and SNIC.  }


\providecommand{\noopsort}[1]{}\providecommand{\singleletter}[1]{#1}%
\appendix

\section{Euler angles}
\label{app:B}
In this appendix we describe how we parameterize the rotation matrix $\ma R$ in terms of Euler angles.
We use Euler angle coordinates in the Goldstein $z$-$x'$-$z''$ convention \cite{Goldstein}: starting from $\ve n^i = \ve e^i$, first rotate the $\ve n^i$ by $\varphi$ around $\ve n^3$, then by $\theta$ around the resulting $\ve n^1$ and finally by $\psi$ around the resulting $\ve n^3$, compare Fig.~\ref{fig:shearplot} in the main text and Fig.~4-7 in Ref.~\cite{Goldstein}. With the shorthand $cx=\cos x$ and $sx=\sin x$ the elements of the rotation matrix are
\begin{align}
	\ma R=\left(
\begin{array}{ccc}
 c\varphi c\psi-c\theta s\varphi s\psi & c\psi s\varphi+c\theta c\varphi s\psi & s\theta s\psi \\
 -c\theta c\psi s\varphi-c\varphi s\psi & c\theta c\varphi c\psi-s\varphi s\psi & c\psi s\theta \\
 s\theta s\varphi & -c\varphi s\theta & c\theta \\
\end{array}
\right)\,.
\end{align}
Evaluating the determinant we can confirm that $\ma R$ is orthogonal.
Our convention is the same as that adopted in Ref.~\cite{Hinch1979}, and Fig.~1 in their paper corresponds to our Fig.~1. Our axis $\ve n^3$  corresponds
to their $z'$-axis.

\section{Quaternion formulation of the Langevin equation, Eq.~(\ref{eqn:langevindef})}\applab{quaternion}
In this appendix we describe how the Langevin equations (\ref{eqn:langevindef}) are expressed in terms of quaternions.
Our quaternion description follows that of \citet{graf2008}. Here we give the practical details relevant for simulation of the Langevin equation (\ref{eqn:langevindef}).
We represent the unit quaternion $q$ as a four-component unit vector $\ve q = (W,X,Y,Z)$, $|\ve q|=1$. Its relation to the rotation matrix $\ma R$ is
\begin{align}
	\ma R = \ma E\cdot \ma G\tr\,,\eqnlab{Rfromq}
\end{align}
where
\begin{equation}
\ma E = \left(
\begin{array}{cccc}
-X & W & -Z & Y \\
-Y & Z & W & -X  \\
-Z & -Y & X & W \\
\end{array}
\right),
\end{equation}
\begin{equation}\label{Gquat}
\ma G = \left(
\begin{array}{cccc}
-X & W & Z & -Y \\
-Y & -Z & W & X  \\
-Z & Y & -X & W \\
\end{array}
\right).
\end{equation}
The equation of motion of $\ve q$ corresponding to Eq.~(\ref{eqn:Rmotion}) is
given in Ref.~\cite{graf2008}:
\begin{equation}
\dot q_i = \tfrac 12 G_{\alpha i} \omega_\alpha\eqnlab{qeqn}\,,
\end{equation}
where $\ve \omega$ is the angular velocity of the particle in body coordinates.
Using Eqs.~(\ref{eq:meanomega}) and (\ref{eqn:omegacorrelation}) we derive
\begin{subequations}
\label{eq:ql}
\begin{align}
\label{eqn:qupdate}
q_i(t+\delta t) & = q_i(t) + \delta q(t)\\
\overline{ \delta q_i } & = \frac{1}{2} G_{\alpha i} \omega^H_\alpha \delta t - \frac{1}{4}\invpe K^{-1}_{\alpha \alpha} q_i \delta t + \mathcal{O}(\delta t ^2),\eqnlab{dqmean}\\
\overline{\delta q_i \delta q_j} & = \frac{1}{2} \invpe G_{\alpha i} K^{-1}_{\alpha \beta} G_{\beta j} \delta t + \mathcal{O}(\delta t ^2)\,.\eqnlab{dqvar}
\end{align}
\end{subequations}
This Langevin equation is equivalent to Eq.~(\ref{eqn:langevindef}) in the main text.

\end{document}